\documentclass[a4paper,11pt]{article}
\pdfoutput=1 
\usepackage[a4paper]{geometry}
\usepackage{amssymb,amsmath,graphicx,amscd,array,color}
\usepackage{amsthm}
\usepackage[round]{natbib}
\usepackage{mathrsfs}
\usepackage{enumerate}
\usepackage{subfigure}    
\usepackage{authblk}
\usepackage{url}
\usepackage{multirow}     

\title{Spatial Modelling of Temperature and Humidity using Systems of Stochastic Partial Differential Equations}
\author[1]{Xiangping Hu\footnote{Corresponding author. Email: \texttt{Xiangping.Hu@math.ntnu.no}}}
\author[1]{Ingelin Steinsland} 
 \author[1]{Daniel Simpson} 
\author[2]{Sara Martino}
\author[1]{H\aa vard Rue} 
\affil[1]{Department of Mathematical Science, Norwegian University of Science and Technology, Trondheim, Norway}
\affil[2]{SINTEF, Trondheim, Norway}

\begin{document}

\maketitle

\begin{abstract} \label{sec: abstract}
In this paper we model spatially temperature and humidity jointly.
The modeling and analysis are based on a dataset for Southern Norway 
which consists of temperature and humidity observations on December 7th each 
year between 2007 and 2011 at about 120 locations. For about half of 
the locations is only temperature available, and not all locations 
are available each year.  
A Bayesian approach is taken, 
and the multivariate Stochastic Partial Differential Equation 
approach for multivariate spatial modeling is used. Hence computationally 
fast inference is available. 
Two different bivariate model as well as an independent model are fitted, 
and the results are in accordance with physical and empirical knowledge. 
The models are further tested and compared with respect to predictability. 
For four out of the five years the bi-variate models are superior the independent model,
especially at locations where only one of the quantities are measured,
 the bivariate model utilize this information for the other quantity.
\end{abstract}

\section{Introduction} \label{sec: spdenorway_introduction}
In many interesting and important situations not only one, but two or more weather variables are important. 
Examples are spring flooding and road maintenance which depends on both precipitation and temperature, and energy demand 
that depends on temperature and wind speed. This paper is motivated by an initiative that aim to develop a 
weather generator that can be used to simulate relevant weather 
variables simultaneously for renewable energy generation and energy demand over larger regions. Instead of 
modeling all variables directly, the strategy is to focus on the variables 
that we know from physics drive the processes. For example it is known that humidity and temperature 
drives precipitation, and these variables are the focus of this paper. 
The strategy is to use a deterministic model to go from temperature and humidity to precipitation. 
Doing so we have a model for precipitation without working with zero-inflated statistical models, and also have a joint model for humidity and temperature.

The modeling are based on a data set of temperature and humidity over Southern Norway for December 7th each year between 2007 and 2011.
A feature of this data set is that humidity and temperature are not necessarily observed at the same locations, and not necessarily each year.
The aim of our work is to build, fit and test a bivariate spatial stochastic model for temperature and humidity for Southern Norway. 
We want to capture both the dependence structure between humidity and temperature as well as their spatial dependencies. 
The models are evaluated on their ability to predict temperature and/or humidity at locations without observations, and at locations where 
the other quantity is observed.

Modelling spatial datasets has been an area of interest for researchers 
in statistics for more then two decades
\citep{cressie1993statistics,stein1999interpolation,diggle2006model,
gelfand2010handbook,cressie2011statistics}. 
Recently \citet{lindgren2011explicit} introduced the Stochastic partial 
differential equations (SPDE) approach to spatial modeling. 
They showed that the SPDE approach coincides with Mat\'ern models.
The motivation to introduce the SPDE-approach was computational as the resulting 
model has Markov properties, and the integrated nested Laplace approximations (INLA) discussed by \citet{rue2009approximate} 
can be used to preform full Bayesian inference. 
But the SPDE-approach also enables 
new modeling opportunities such as oscillating dependency structure \citep{lindgren2011explicit}, 
non-stationary models with explanatory variables in the dependency model \citep{ingebrigtsen2012using} 
and non-stationary models driven by vector fields \citep{fuglstad2014exploring}. 
Nested SPDEs were proposed by \citet{bolin2011spatial} for constructing a larger class of models for spatial datasets. 

When more then one response variable is of interest, we need to use multivariate models. 
Spatial multivariate modeling has been used for models in economics \citep{gelfand2004nonstationary, sain2007spatial}, 
in the area of air quality \citep{brown1994multivariate, schmidt2003bayesian}, weather forecasting 
\citep{courtier1998ecmwf,reich2007multivariate} and quantitative genetics \citep{mcguigan2006studying, konigsberg2009multivariate}.
For multivariate spatial phenomenons several approaches have been proposed, 
such as linear model of coregionalization (LMC) \citep{goulard1992linear, wackernagel2003multivariate, gel2004calibrated} and 
covariance-based models \citep{apanasovich2010cross, gneitingmatern,li2011approach,kleiber2012nonstationary,apanasovich2012valid}.
\citet[Chapter 3.12.3]{ribeiromodel} pointed out that models constructed with LMC approach usually are poorly identifiable without
some restrictions being placed beforehand on the processes. A challenge for the covariance based models is to construct a  
positive definite matrix. The computational burden for these models are also very high
due to the cost of $\mathcal{O}(n^3)$ to factorize a dense $n \times n$ covariance matrix.
Recently, \citet{hu2012multivariate, hu2012oscillating} introduced multivariate SPDE-models. 
This SPDE approach has both the computational benefit of the univariate SPDE models, and are by construction positive definite.

In this paper we use the multivariate SPDE models discussed by \citet{hu2012multivariate} to model temperature and humidity in Southern Norway. 
The models we propose have year specific intercepts and spatial fields, but the spatial fields are considered replicates 
of the same spatial process as they share spatial parameters. 

The rest of this paper is organized as followings. Section \ref{sec: spdenorway_data} describes the data.  
We review the knowledge about the SPDE approach for spatial statistics in Section \ref{sec: spdenorway_background}. 
Section \ref{sec: spdenorway_model} describes the spatial model for our dataset.
Section \ref{sec: spdenorway_evaluation} discusses the evaluation procedure. 
Results are given in Section \ref{sec: spdenorway_results}. 
Section \ref{sec: spdenorway_discussion} ends the paper with discussion and conclusion.

\section{Temperature and Humidity in Southern Norway} \label{sec: spdenorway_data}
\begin{figure}[tbp]
    \centering
     \subfigure[]{\includegraphics[width=0.45\textwidth,height=0.45\textwidth]{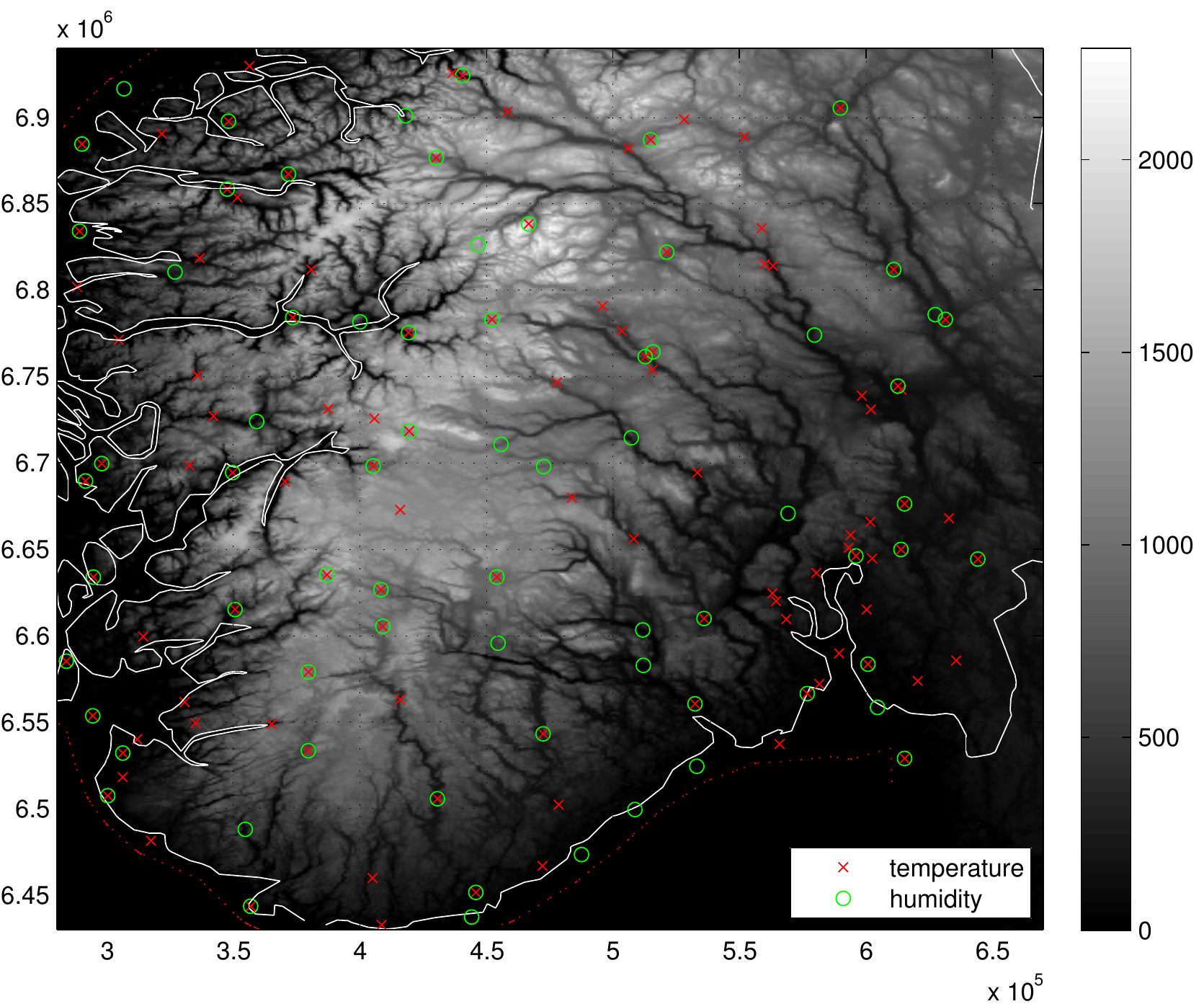} \label{fig: spdenorway_covariates_elevation} } 
     \subfigure[]{\includegraphics[width=0.45\textwidth,height=0.45\textwidth]{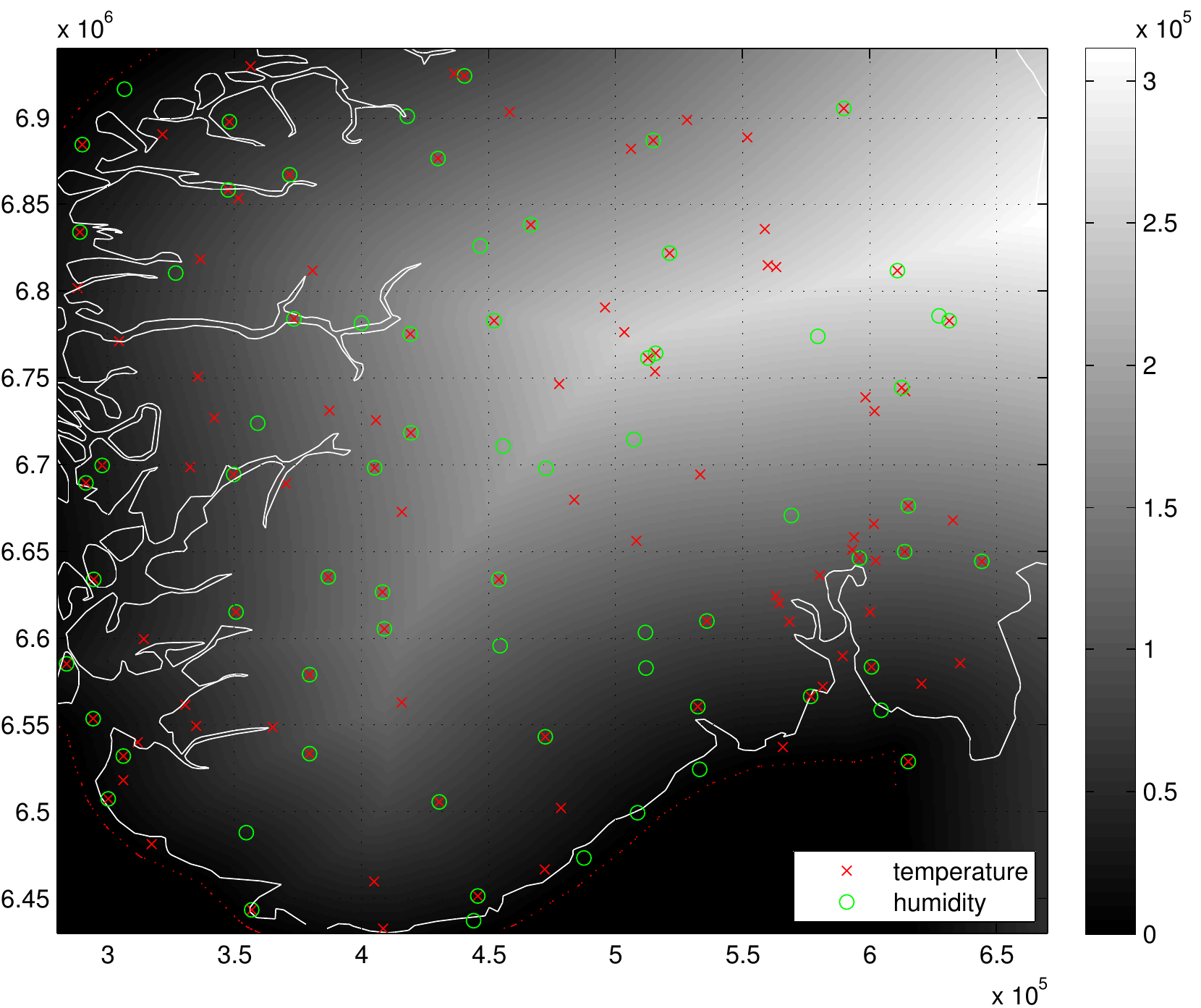}  \label{fig: spdenorway_covariates_distance} }
    \caption{Locations of temperature and humidity observations on $7$th of December in $2011$ with elevation 
             (a) and distances to ocean (b) on a $1$km by $1$km grid.
             The base line for calculating the distance to ocean (dotted-line) and the coast line (solid line) of southern Norway are also given. 
             The cross marks ($\times$) and the circle marks ($\circ$) are locations for temperature and humidity observations, respectively. Both covariates are in meters.} 
\label{fig: norway_data_with_covariates}
 \end{figure}

We build the analysis in this paper on a dataset containing observations for temperature
and humidity on $7$th of December each year from $2007$ to year $2011$, i.e. for $5$ years. 
The temperature dataset contains daily mean temperature in Celsius degree and the humidity dataset contains the measured mixing ratio of humidity.
The mixing ratio of humidity is defined as the mass of water vapor contained in a unit mass of dry air, and hence has a unit kg/kg. 
It is important to point it out that the observations are not necessarily at the same locations for all the $5$ years.
Most of humidity observations are measured at a subset of locations of temperature. 
Two covariates are used in the model: elevation at the measurement location and the distance to the ocean.
Figure \ref{fig: spdenorway_covariates_elevation} and Figure \ref{fig: spdenorway_covariates_distance}
give an overview of locations for temperature and humidity. 
The dotted line is the base line for calculating the distance to ocean and the solid line is the 
coast line of southern Norway. We can clearly note that the distance to ocean is not the same as the distance to the coast. 

\section{Background}  \label{sec: spdenorway_background}
\subsection{Univariate GRFs in SPDE formulation} \label{sec: spdenorway_univariateSPDE}

The main idea of the newly proposed approach by \citet{lindgren2011explicit} is to use an SPDE to construct GRFs for modelling spatial datasets.  
The SPDE used in this paper has the form
\begin{equation} \label{eq: spdenorway_spde_simple}
 b (\kappa^2 - \Delta)^{\alpha/2} {x}(\boldsymbol{s}) = \mathcal{W}(\boldsymbol{s}), \hspace{3mm} \boldsymbol{s} \in \mathbb{R}^d, \hspace{3mm} \alpha = \nu +d/2, \hspace{3mm}  \nu >0,
\end{equation}
 where $b$ is a parameter related to the variance of the random field $x(\boldsymbol{s})$, $\mathcal{W}(\boldsymbol{s})$ 
 is a standard Gaussian white noise process,
 $(\kappa^2 - \Delta)^{\alpha/2}$ is a pseudo (fractional) differential operator and $\alpha$ must be a non-negative integer. 
$\Delta$ is the standard \emph{Laplacian} with definition
\begin{displaymath}
\Delta = \sum_{i = 1}^{d}{\frac{\partial^2}{\partial x_i^2}}.
\end{displaymath}
\citet{whittle1954stationary, whittle1963stochastic} has shown that the stationary solution $x(\boldsymbol{s})$ to the SPDE \eqref{eq: spdenorway_spde_simple} is a GRF with a Mat\'ern covariance function. 
The Mat\'ern covariance function has the form
 \begin{equation} \label{eq: spdenorway_matern_covariance}
M(\boldsymbol{h} |\nu, \kappa) = \frac{\sigma^2 2^{1-\nu}}{\Gamma(\nu)}(\kappa \| \boldsymbol{h} \|)^\nu K_\nu(\kappa \| \boldsymbol{h} \|),
\end{equation}
where $\nu$ is the smoothness parameter, $\kappa$ is the scaling parameter and $K_{\nu}$ is the modified Bessel function of second kind with order $\nu$, 
$\|h\|$ denotes the Euclidean distance in $\mathbb{R}^d$
 and $\sigma^2$ is the marginal variance. The Mat\'ern covariance function is isotropic and it is widely used in spatial statistics \citep{stein1999interpolation, 
diggle2006model,simpson2010order,lindgren2011explicit,bolin2011spatial,ingebrigtsen2012using, 
hu2012multivariate, hu2012oscillating}. 
In this approach we use the finite element methods (FEMs) to solve the SPDE \eqref{eq: spdenorway_spde_simple}, and then apply the GMRF approximation 
to the solution in order to obtain computationally efficient inference. 
\citet{bolin2009wavelet} showed that the differences between the exact FEM representation and the GMRFs approximation are negligible. 
Since the smoothness parameter $\nu$ is poorly identifiable \citep{diggle2006model, lindgren2011explicit}, 
we fix $\alpha$ for all our models to $\alpha = 2$.

\subsection{Multivariate GRFs in SPDE formulation} \label{sec: spdenorway_multivariateSPDE}
\citet{hu2012multivariate} have extended the approaches from \citet{lindgren2011explicit} to construct multivariate GRFs. 
This approach for constructing multivariate GRFs inherits both theoretical
and computational advantages from the approach given by \citet{lindgren2011explicit} for univariate GRFs. 
The system of SPDEs for constructing a $p$-dimensional multivariate GRF has the form 
\begin{align} \label{eq: spdenorway_SPDEs_system}
\begin{pmatrix}
\mathcal{L}_{11} & \mathcal{L}_{12} & \ldots & \mathcal{L}_{1p}\\
\mathcal{L}_{21} & \mathcal{L}_{22} & \ldots & \mathcal{L}_{2p}\\
\vdots & \vdots & \ddots & \vdots\\
\mathcal{L}_{p1} & \mathcal{L}_{p2} &\ldots & \mathcal{L}_{pp}
\end{pmatrix}
\begin{pmatrix}
x_1(\boldsymbol{s})\\ x_2(\boldsymbol{s})\\ \vdots \\ x_p(\boldsymbol{s})
\end{pmatrix}
=
\begin{pmatrix}
\varepsilon_1(\boldsymbol{s}) \\ \varepsilon_2(\boldsymbol{s}) \\ \vdots \\ \varepsilon_p(\boldsymbol{s})
\end{pmatrix},
\end{align}
where $\mathcal{L}_{ij} = b_{ij}(\kappa_{ij}^2 - \Delta)^{\alpha_{ij}/2}$ are similar differential operators as 
given in Equation \eqref{eq: spdenorway_spde_simple}
with $\{ \alpha_{ij} = 0 \text{ or } 2; 1 \leq i, j \leq p \}$, $\{ \varepsilon_i(\boldsymbol{s}); i, j = 1,\dots,p \}$
are Gaussian noise processes which are independent but not necessarily identically distributed. It was shown by \citet{hu2012multivariate} that the solution 
$\boldsymbol{x}(\boldsymbol{s}) = \left( x_1(\boldsymbol{s}), x_2(\boldsymbol{s}), \dots, x_p(\boldsymbol{s}) \right)$ 
to the system of SPDE \eqref{eq: spdenorway_SPDEs_system}
 is a multivariate GRF. The parameters
$\{ \kappa_{ij}; i, j = 1,\dots,p \}$ and $\{ \nu_{ij}; i, j = 1,\dots,p\}$ are scaling parameters and smoothness parameters, respectively. 
$\{b_{ij}; i, j = 1,\dots,p \}$ are related to both the marginal variances of the fields and the cross covariances among the GRFs. 
Further, similarly as discussed by \citet{lindgren2011explicit}, the precision matrix $\boldsymbol{Q}$ (inverse of the covariance matrix)
for the multivariate GRF constructed from the system of SPDEs \eqref{eq: spdenorway_SPDEs_system} satisfies the positive definite constraint automatically. 
\citet{hu2012multivariate} demonstrated that the link between the GMRFs and GRFs could be used, 
and hence we can construct models with GRFs but use GMRFs for computations. Since the precision 
matrix $\boldsymbol{Q}$ of the multivariate GMRF $\boldsymbol{x}(\boldsymbol{s})$ is sparse.
Therefore numerical algorithms for sparse matrices can be applied for fast sampling and inference. 
 
We follow \citet{hu2012multivariate, hu2012oscillating} and use a triangular system of SPDEs
\begin{align} \label{eq: spdenorway_SPDEs_system_triangular}
\begin{pmatrix}
\mathcal{L}_{11} &   \\
\mathcal{L}_{21} & \mathcal{L}_{22} \\
\end{pmatrix}
\begin{pmatrix}
{x}_1(\boldsymbol{s}) \\ {x}_2(\boldsymbol{s})
\end{pmatrix}
=
\begin{pmatrix}
\mathcal{W}_1(\boldsymbol{s}) \\ \mathcal{W}_2(\boldsymbol{s})
\end{pmatrix},
\end{align}
where $\{ \mathcal{W}_i(\boldsymbol{s}); i = 1,2 \}$ are standard Gaussian white noise processes.
This is a special case of the system of Equations \eqref{eq: spdenorway_SPDEs_system} with 
$\mathcal{L}_{12} = 0$ and $\{ \varepsilon_i(\boldsymbol{s}) = \mathcal{W}_i(\boldsymbol{s}); i = 1,2 \}$ when $p = 2$.
The advantage of a triangular systems of SPDEs is that this simplification makes both computations and interpretation easier.

With this setting we know that $x_1(\boldsymbol{s})$ is a Mat\'ern random field and 
$x_2(\boldsymbol{s})$ is generally not a Mat\'ern random field, but close to a Mat\'ern random field \citep{hu2012multivariate}.
This implies that the order of the random fields matters.
Generally speaking, we need to choose the order of the random fields $x_1(\boldsymbol{s})$ and $x_2(\boldsymbol{s})$,
and this is usually done by a model selection test. Fit models with both orders and pick the one that minimizes some criterion, such as prediction error. 
Using the triangular system of SPDEs \eqref{eq: spdenorway_SPDEs_system_triangular} for constructing a bivariate GRF, 
we have $6$ parameters to estimate $\boldsymbol{\theta} = \left\{ \kappa_{11}, \kappa_{21}, \kappa_{22}, b_{11}, b_{21}, b_{22} \right\}$ 
from the system of SPDEs when we model the temperature and humidity jointly.

\citet{hu2012multivariate} showed that the sign of cross-correlation between $x_1(\boldsymbol{s})$ and $x_2(\boldsymbol{s})$ is only related to
the product $b_{21}b_{22}$ with a triangular system of SPDE. In the extreme case, if $b_{21}$ is zero, i.e., 
$x_1(\boldsymbol{s})$ and $x_2(\boldsymbol{s})$ are independent, 
then $b_{22}$ can only be positive value. Therefore we restrict $b_{22}$ to be positive and then 
the sign of the cross-correlation is decided by the sign of $b_{21}$. 
When $b_{21} < 0$, $x_1(\boldsymbol{s})$ and $x_2(\boldsymbol{s})$ are positively correlated, and when $b_{21} > 0$, 
$x_1(\boldsymbol{s})$ and $x_2(\boldsymbol{s})$ are negatively correlated.

\section{Models for temperature and humidity} \label{sec: spdenorway_model}
We now set up three different Bayesian hierarchical models for temperature and humidity. 
The models have three levels, data model, process model and parameter models.
We propose models differ in the dependency between temperature and humidity, i.e. they have different process models. 
To do inference we have available data that we denote $y_{ijk}$, where $i$ is a location index,
$j$ a year index and $k$ a field index. We have $K=2$ fields, $k \in \{ T, H \}$, 
temperature and humidity, respectively. Further we have available data for $J=5$ years, 
$j \in \{ 2007, 2008, \dots, 2011 \}$. For each field $k$ and each year $j$ there are available 
observations at $N_{jk}$ locations (see Table \ref{tab: norway_data_numbersobservation}), $i \in \{1,2,\dots N_{jk} \}$,
and the observations are not necessary measured at the same locations in each year.

\begin{table}
\caption{\label{tab: norway_data_numbersobservation} Number of observations for temperature and humidity}
    \centering
\fbox{%
 \begin{tabular}{*{6}{c}}
\hline
    Year       &  $2007$   &  $2008$    &     $2009$    &   $2010$    &    $2011$       \\
\hline
 temperature   &   $97$    &   $104$    &     $111$     &   $122$     &    $128$        \\
  humidity      &   $56$    &   $63$     &     $62$      &   $62$     &    $70$          \\
\hline
 \end{tabular}}
\end{table}

All models have the same data model. Given the truth $\eta_{ijk}$ the observed data is assumed to be Gaussian, 
\begin{equation}
y_{ijk}| \eta_{ijk} \sim \mathcal{N}(\eta_{ijk}, \sigma_k^2)
\end{equation}
for $k \in \{T, H\}$. The variances $\sigma_{T}^2$ and $\sigma_H^2$ are interpreted to come from measurement uncertainty 
for temperature and humidity respectively. From knowledge about the measurement process we set the variances to $\sigma_T^2= 0.1^2$ and $\sigma_H^2=0.01^2$. 

For each year $j$ the true temperature field ${\bf{\eta}}_{jT}$ and humidity field ${\bf{\eta}}_{jT}$ are modeled as a 
linear combination of explanatory variables and a spatial field. We use three explanatory variables, year as a factor, 
${\bf{\beta}}_{year,k} = (\beta_{2007, k}, \beta_{2008,k}, \dots, \beta_{2011,k})$ 
and elevation ($\beta_{evla,k}$) and distance to ocean ($\beta_{dist,k}$) as linear effects. 
The model can be written in vector form as:
\begin{equation}
{\bf{\eta}}_{jT} = {\bf{\beta}}_{T} X_j + {\bf{\xi}}_{jT},
\end{equation}  
\begin{equation}
{\bf{\eta}}_{jH} = {\bf{\beta}}_{H} X_j + {\bf{\xi}}_{jH},
\end{equation}
where ${\bf{\beta}}_{T} = ({\bf{\beta}}_{year,T}, \beta_{evla,T}, \beta_{dist,T})$,
${\bf{\beta}}_{H} = ({\bf{\beta}}_{year,H}, \beta_{evla,H}, \beta_{dist,H})$, $X_j$ are design matrices
and ${\bf{\xi}}_{jT}$ and ${\bf{\xi}}_{jH}$ are spatial fields for year $j$ for temperature and
humidity, respectively. We now set up three models for the spatial fields 
$\xi_j = (\xi_{jT}, \eta_{jH})$ which are all SPDE models as introduced in Section \ref{sec: spdenorway_univariateSPDE} and \ref{sec: spdenorway_multivariateSPDE}.  
Our first model assumes independent univariate SPDE models (UM) for $\xi_{jT}$ and $\xi_{jH}$.
The other two models allow for dependent fields, and are bivariate SPDE models. 
The triangular system of SPDEs in Section \ref{sec: spdenorway_multivariateSPDE} gives us two modeling opportunities regarding the ordering: 
Modeling temperature as the first field (i.e. as a Mat\'ern field), and humidity as the second field, 
which we denote BM-TH; or to model humidity as the first field and temperature as the second,
BM-HT. Further, we assume that the fields $\eta_j$ for the different years $j$ are independent realizations of the same models, 
i.e. the parameters of the SPDE models do not change from year to year.
The models for $\xi_j= (\xi_{jT}, \xi_{jH})$ are summarized below:

\begin{description}
\item[Independent univariate SPDE model (UM):] $\xi_{jT}$ and $\xi_{jH}$ are assumed independent and
\begin{equation}
\xi_{jT} \sim UM(b_T, \kappa_T)
\end{equation}
\begin{equation}
\xi_{jH} \sim UM(b_H, \kappa_H)
\end{equation}
\item[Bivariate SPDE model, TH (BM-TH):] We model temperature as the first field and humidity as the second.
$\xi_{jT}$ and $\xi_{jH}$ are assumed to follow a bivariate SPDE-model 
with temperature as the first field in the formulation in Equation \eqref{eq: spdenorway_SPDEs_system_triangular};
\begin{equation}
\xi_{j} \sim BM ({\boldsymbol{b}}_{TH}, {\boldsymbol{\kappa}}_{TH})
\end{equation}
where ${\boldsymbol{b}}_{TH} = (b_{T}, b_{HT}, b_{HH})$ and ${\boldsymbol{\kappa}}_{TH}= (\kappa_{T}, \kappa_{HT},\kappa_{HH})$. 
\item[Bivariate SPDE model, HT (BM-HT):] We model humidity as the first field and temperature as the second.
$\xi_{jT}$ and $\xi_{jH}$ are assumed to follow a bivariate SPDE-model 
with himidity as the first field in the formulation in Equation \eqref{eq: spdenorway_SPDEs_system_triangular};
\begin{equation}
\xi_{j} \sim BM({\bf b}_{HT}, {\boldsymbol{\kappa}}_{HT})
\end{equation}
where ${\boldsymbol{b}}_{HT} \sim (b_{H}, b_{TH}, b_{TT})$ and ${\boldsymbol{\kappa}}_{HT}= (\kappa_{H}, \kappa_{TH},\kappa_{TT})$. 
\end{description}
 
The model formulations are completed by assigning priors to the parameters. 
All explanatory variable parameters ($\beta$s) are given independent vague Gaussian priors;
$\beta \sim N(0, 100)$, while the SPDE parameters ($\kappa$s and $b$s) are given 
independent log-Gaussian priors: $\log \kappa \sim N(0, 100)$ and $\log b \sim N(0, 100)$.

\section{Evaluation} \label{sec: spdenorway_evaluation}

In this section we describe the scores and evaluation schemes used to compare the results of three different models 
set up in Section \ref{sec: spdenorway_model}.

\subsection{Scoring rules} \label{sec: spdenorway_scoringrules} 
In this paper the commonly used scoring rules mean absolute error (MAE), mean-square error (MSE) and the average of the continuous ranked probability score (CRPS) are chosen. 
Let $\hat{y}_{ijk}$ denote the prediction for the observations $y_{ijk}$ for 
the observation $i$ in year $j$ for the $k$th field, and then the MAE and MSE for the $k$th field have the following definitions
\begin{equation*} \label{eq: spdenorway_MAEMSE}
 \begin{split}
   \text{MAE}_k & = \frac{1}{n_k}\sum_{j}\sum_{i}|y_{ijk}-\hat{y}_{ijk}|, \\
   \text{MSE}_k & =  \frac{1}{n_k}\sum_{j}\sum_{i}(y_{ijk}-\hat{y}_{ijk})^2, \\
 \end{split}
\end{equation*}
The CRPS is also a commonly used scoring rule to evaluate the probabilistic forecasts, and it is the integral of the 
Brier scores for a continuous predictand at all possible threshold values $p$ \citep{hersbach2000decomposition,gneiting2005calibrated}.
Let $F$ denote the predictive cumulative distribution function (CDF) and 
$H(p-y)$ be the Heaviside function with value $1$ whenever $p-y>0$ and value 0 otherwise. Then the continuous ranked probability score is defined as 
\begin{equation}
 \text{crps}(F,y)  = \int_{-\infty}^{\infty}{\left(F(p)-H(p-y)\right)^2}dp.
\end{equation}
\citet{gneiting2005calibrated} pointed out that if $F$ is the CDF of a Gaussian distribution, 
then a closed form of the continuous ranked probability score can be obtained, and this form is usually used in applications. 
The average of continuous ranked probability score, CRPS, then has the form
\begin{equation}
\text{CRPS}_k = \frac{1}{n_k}\sum_j \sum_i{\text{crps}(F_{ijk}, y_{ijk})}.
\end{equation}

\subsection{Validation scheme} \label{sec: spdenorway_crossvalidation}
To evaluate the predictive performance we use validation scheme where the data set is divided into a training set and a test set. 
The test sets consist of 20 locations that are chosen at random for each year among the locations that have observations for both temperature and humidity that year. 
The same test set is used for all three models and the following validation scheme has been chosen for comparing the results from different models. 

\textbf{Setting H} \newline
\indent In this setting only predictive performance for humidity is evaluated. The model is fitted by using all 
data but humidity observations for the test locations.

\textbf{Setting T} \newline
\indent In this setting only predictive performance for temperature is evaluated. The model is fitted by using all 
data but temperature observations for the test locations.

\textbf{Setting HT} \newline
\indent In this setting both the predictive performance for temperature and humidity is evaluated. The model is fitted by using all 
data but temperature and humidity observations for the test locations.

\section{Results} \label{sec: spdenorway_results}
In this seciton some empirical data analysis have been conducted in Section \ref{sec: spdenorway_empiricalanalysis}. Inference results of the parameters
of the models set up in Section \ref{sec: spdenorway_model} are given in Section \ref{sec: spdenorway_bayesianparameters}, 
while the results for predictive performance are given in Section \ref{sec: spdenorway_predictiveperformance}.

\subsection{Empirical data analysis} \label{sec: spdenorway_empiricalanalysis}
Since the numerical values of humidity observations are positive, they are preprocessed with the widely used Box-Cox family of transformations
in order to transform them to be approximately Gaussian distributed \citep{box1964analysis}. 
The Box-Cox family of transformations has the form 
\begin{equation}
 \hat{Y} = \begin{cases}
             \left( Y^\lambda - 1 \right)/\lambda   & \text{if } \lambda \neq 0  \\
             \log(Y)                              & \text{if }  \lambda = 0
           \end{cases}.
\end{equation}
The estimated value of $\lambda$ for the Box-Cox transform is $\lambda = 0.66$. The transformation function is a monotonic increase function and the transformed humidity
is more reasonable to be modelled with Gaussian distribution. We use the original observations of temperature.
\citet{sakia1992box} and \citet{diggle2006model} give more information about the Box-Cox transformation and other transformation methods. 

The empirical variograms of both temperature and humidity have been calculated and fitted to theoretical variograms. 
In the theoretical variograms, we choose to fit with the Mat\'ern model. This analysis suggest the smoothness parameters for both the fields with $\nu = 1$ are reasonable, 
and hence fixing $\alpha = 2$ in our analysis is also reasonable. 

\subsection{Inference results of parameters} \label{sec: spdenorway_bayesianparameters}
We follow \citet{rue2009approximate} and treat the coefficients for the covariates, and the yearly effects of 
temperature and humidity as parts of the latent field $\boldsymbol{z} $, i.e.,
 $\boldsymbol{z} = \left( \boldsymbol{x}, \boldsymbol{\beta} \right)^{\mbox{T}}$, to achieve computational efficiency. This is due to the fact that there will be 
much fewer parameters during the optimization. Detailed setting of inference is given in {Appendix B}. 
It can be shown that
\begin{equation} \label{eq: spdenorway_x|y,theta}
\begin{split}
 \pi(\boldsymbol{z}|\boldsymbol{y}, \boldsymbol{\theta}) & \propto \pi({\boldsymbol{z}, \boldsymbol{y} | \boldsymbol{\theta}})   \\
          & = \pi(\boldsymbol{z}|\boldsymbol{\theta}) \pi(\boldsymbol{y}|\boldsymbol{z}, \boldsymbol{\theta}) \\
          & \propto \exp \left( -\frac{1}{2} \left[ \boldsymbol{z}^{\mbox{T}} (\boldsymbol{Q}(\boldsymbol{\theta}) + 
           \boldsymbol{C}^{\mbox{T}} \boldsymbol{Q}_n \boldsymbol{C}) \boldsymbol{z} - 2\boldsymbol{z}^{\mbox{T}} \boldsymbol{C}^{\mbox{T}}\boldsymbol{Q}_n \boldsymbol{y} \right] \right),
\end{split}
\end{equation}
and
\begin{equation} \label{eq: spdenorway_x|y,theta_canonical}
 {\boldsymbol{z}|\boldsymbol{y}, \boldsymbol{\theta}} \sim \mathcal{N} \left( \boldsymbol{\mu}_c (\boldsymbol{\theta}), \boldsymbol{Q}_c (\boldsymbol{\theta}) \right),
\end{equation}
with $\boldsymbol{\mu}_c$, $\boldsymbol{Q}_c$ and $\boldsymbol{C}$ given in {Appendix B}. 
From Equation \eqref{eq: spdenorway_x|y,theta} we can get the estimates for the yearly effects and for the coefficients of the covariates.
For model BM-TH, we set $\boldsymbol{x}_1$ as temperature and $\boldsymbol{x}_2$ as humidity,
and the estimates for the yearly effects are given in Table \ref{tab: spdenorway_estimate_TempHumi_yeareffects}, with standard deviations given in brackets. 
Table \ref{tab: spdenorway_estimate_TempHumi_yeareffects} shows that the yearly effects are quite different. 
This explains the high temperature in $2007$ but low temperature in $2010$. The estimates of the coefficients of the covariates are 
given in Table \ref{tab: spdenorway_TempHumi_covariatescoffcients}. We can notice that 
the two covariates give negative contribution to both fields. Similar results can be obtained when we change the order of the fields for the systems of 
SPDEs given in Equation \eqref{eq: spdenorway_SPDEs_system_triangular},
i.e., when we set the first field $\boldsymbol{x}_1$ as humidity and the second field $\boldsymbol{x}_2$ as temperature.
These results agree with physical knowledge, and we summarize as follows:
the higher elevation, the lower temperature; the higher elevation, the lower humidity; the longer distance to ocean, the lower temperature, and
the longer distance to ocean, the lower humidity. We have standardized the elevation and distance to ocean by divide them with $2 \times 10^3$ and 
$2.5 \times 10^5$, respectively, and hence the inference is more stable.

\begin{table}
 \caption{\label{tab: spdenorway_estimate_TempHumi_yeareffects} Posterior modes for yearly effects for different years with bivariate models and univariate model}
  \centering
\fbox{%
 \begin{tabular}{*{7}{c}}
   Model                     &    Parameter                       &  $2007$   &   $2008$  &   $2009$   &   $2010$   &   $2011$    \\
\hline
 \multirow{4}{*}{BM}         &  \multirow{2}{*}{$\beta_{year,T}$}  & $8.81$   & $-0.03$   & $8.57$    & $-5.98$   & $0.89$   \\
                             &                                 & ($ 0.55$) & ($0.54$) & ($0.53$)  & ($0.53$)  & ($0.52$) \\
                             &  \multirow{2}{*}{$\beta_{year,H}$}  & $3.04$   & $1.82$    & $2.86$    & $1.15$    & $1.67$    \\
                             &                                 & ($0.11$) & ($0.11$)  & ($0.11$)  & ($0.11$)  & ($0.11$)  \\
\hline
 \multirow{4}{*}{UM}         &  \multirow{2}{*}{$\beta_{year,T}$}  & $8.79$   & $-0.05$   & $8.54$    & $-6.00$   & $0.87 $     \\
                             &                                 & ($0.55$) & ($0.54$)  & ($0.53$)  & ($0.53$)  & ($0.52$)    \\
                             &  \multirow{2}{*}{$\beta_{year,H}$}  & $3.08$   & $1.84$    & $2.90$    & $1.17$    & $1.70$      \\
                             &                                 & ($0.12$) & ($0.12$)  & ($0.12$)  & ($0.12$)  & ($0.12$)    \\
 \end{tabular}}
\end{table}

\begin{table}
 \caption{\label{tab: spdenorway_TempHumi_covariatescoffcients} Posterior modes and standard deviations of coefficients for covariates}
 \centering
\fbox{%
 \begin{tabular}{*{4}{c}}
    Model                      & Parameter      & Estimate     &    Std. dev.  \\
  \hline
 \multirow{4}{*}{BM}          & $\beta_{elevation, T}$   & $-6.83$     &  $0.68$      \\
                              & $\beta_{distance, T}$   & $-9.86$     &  $0.72$      \\
                              & $\beta_{elevation, H}$   & $-0.20$     &  $0.09$       \\
                              & $\beta_{distance, H}$   & $-1.46$     &  $0.13$       \\
 \hline
 \multirow{4}{*}{UM}         & $\beta_{elevation, T}$   & $-6.83$    &  $0.68$       \\
                             & $\beta_{distance, T}$   & $-9.82$    &  $0.72$       \\
                             & $\beta_{elevation, H}$   & $-0.49$    &  $0.11$       \\
                             & $\beta_{distance, H}$   & $-1.44$    &  $0.14$       \\
 \end{tabular}}
\end{table}

\begin{figure}[tbp]
    \centering
    \mbox{\includegraphics[width=0.8\textwidth,height=0.5\textwidth]{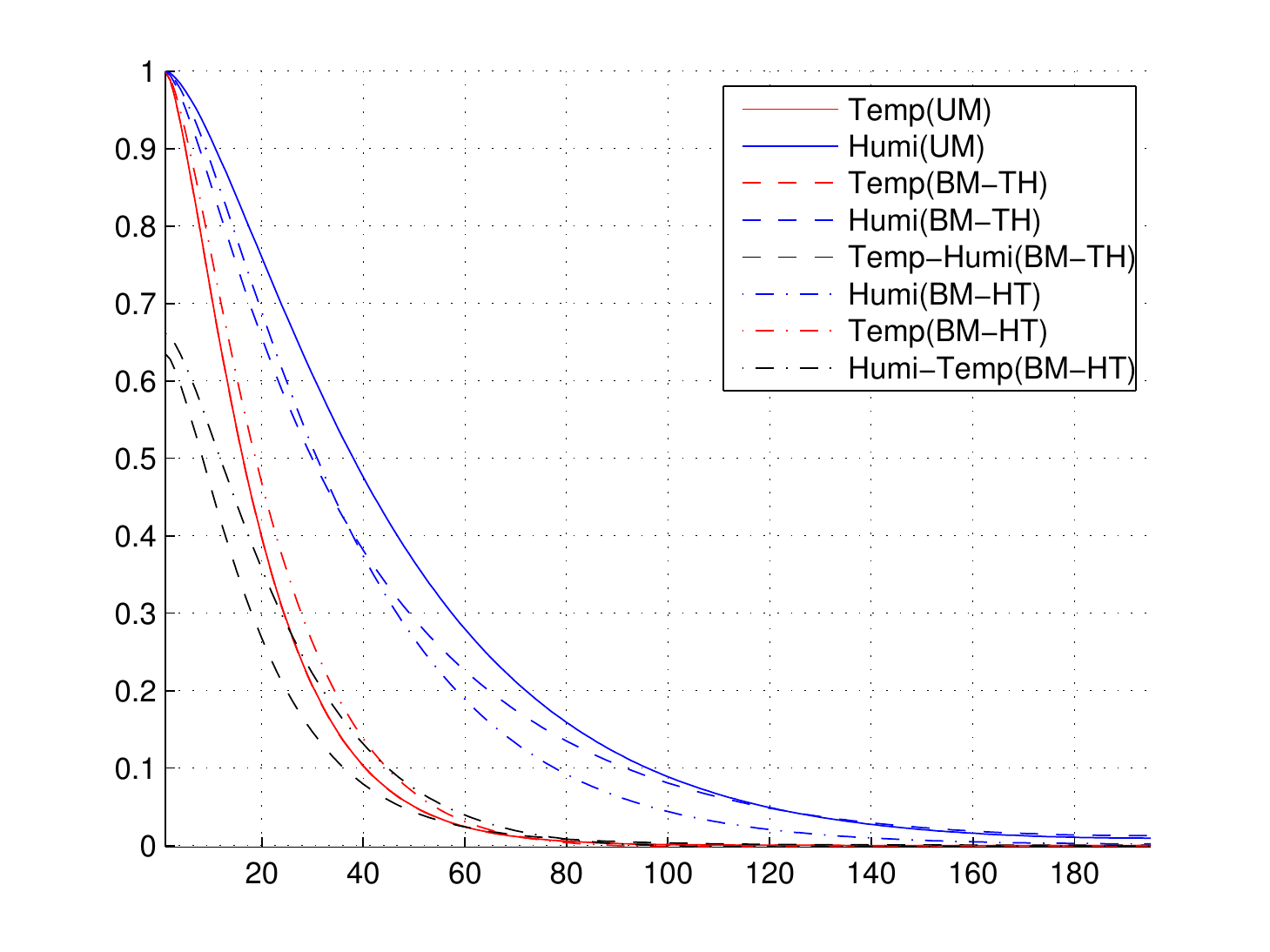}}
    \caption{\label{fig: spdenorway_correlation_BMUM}Correlations within temperature and humidity and cross-correlation 
             between temperature and humidity for bivariate model and univariate model. The correlation structures for temperature and humidity are shown in red and blue colors, respectively,
             and the cross-correlation between temperature and humidity are shown with black colors. Different properties of lines indicate different models,
             i.e., the solid, dashed and dash-dot lines indicate the results from UM, BM-TH and BM-HT, respectively.} 
\end{figure}

The posterior mean estimates estimates and the posterior standard deviations are given for the bivariate models
in Table \ref{tab: spdenorway_estimate_spde}. We notice that temperature and humidity are positively 
correlated since $b_{HT} < 0$ and $b_{TH} < 0$ for the two models.
The results for all three models, i.e., UM, BT-TH and BT-HT, are given in Figure \ref{fig: spdenorway_correlation_BMUM}.  

From the results shown in Table \ref{tab: spdenorway_correlationrange} and \ref{fig: spdenorway_correlation_BMUM} 
we notice that the correlation range differ between models. We further notice that
the correlation ranges of humidity and temperature from UM are the longest and shortest, respectively comparing to the results from BM-TH and BM-HT.
The cross-correlations between temperature and humidity at the same location are $\gamma = 0.64$ and $\gamma = 0.66$ for BM-TH and BM-HT, respectively. 
From these results we conclude that the cross-correlation between temperature and humidity are relatively high and indeed needed to be considered. 

\begin{table}
 \caption{\label{tab: spdenorway_estimate_spde} Posterior modes for hyper-parameters of bivariate model and of univariate model}
  \centering
\fbox{%
  \begin{tabular}{*{4}{c}}
             & BM-TH                                    &   BM-HT                               &            UM    \\
\hline
 $b_{TT}$  & $1.04 \times 10^{-2}$  ($8.106 \times 10^{-4}$) &  $1.98 \times 10^{-2}$  ($2.843 \times 10^{-3}$) &  $1.04 \times 10^{-2}$ ($8.134 \times 10^{-4}$) \\
 $b_{TH}$  & $-2.19\times 10^{-2}$ ($2.512 \times 10^{-3}$)  &  $-2.23 \times 10^{-1}$ ($2.875 \times 10^{-2}$) &                                              \\
 $b_{HH}$  & $3.13 \times 10^{-1}$  ($2.04 \times 10^{-2}$)  &  $1.71 \times 10^{-1}$  ($1.79 \times 10^{-2}$)  &  $2.15\times 10^{-1}$ ($1.460 \times 10^{-2}$) \\
 $\kappa_{TT}$            & $7.69$  ($0.64$)                     &   $5.64$  ($0.72$)                &      $7.67$ ($0.64$)                         \\
 $\kappa_{TH}$            & $3.23$ ($0.60$)                      &   $2.54$  ($0.59$)                &                                                  \\
 $\kappa_{HH}$            & $2.80$ ($0.40$)                      &   $3.95$  ($0.41$)              &      $3.20$ ($0.27$)                         \\
 \end{tabular}}
\end{table}

\begin{table}
\caption{\label{tab: spdenorway_correlationrange}Correlation ranges for bivariate models (BM-TH and BM-HT) and univariate model (UM)}
\centering
\fbox{%
  \begin{tabular}{*{4}{c}}
                     & $\rho_T$   & $\rho_H$              &  $\rho_{TH}$        \\
\hline
 BM-TH    & $39.4$km   & $90.7$km              &  $35.2$km             \\
 BM-HT    & $43.7$km   & $76.7$km              &  $43.8$km             \\
 UM       & $39.4$km   & $94.9$km              &                     \\
 \end{tabular}}
\end{table}

\begin{figure}[tbp]
    \centering
     \subfigure[]{\includegraphics[width=0.3\textwidth,height=0.3\textwidth]{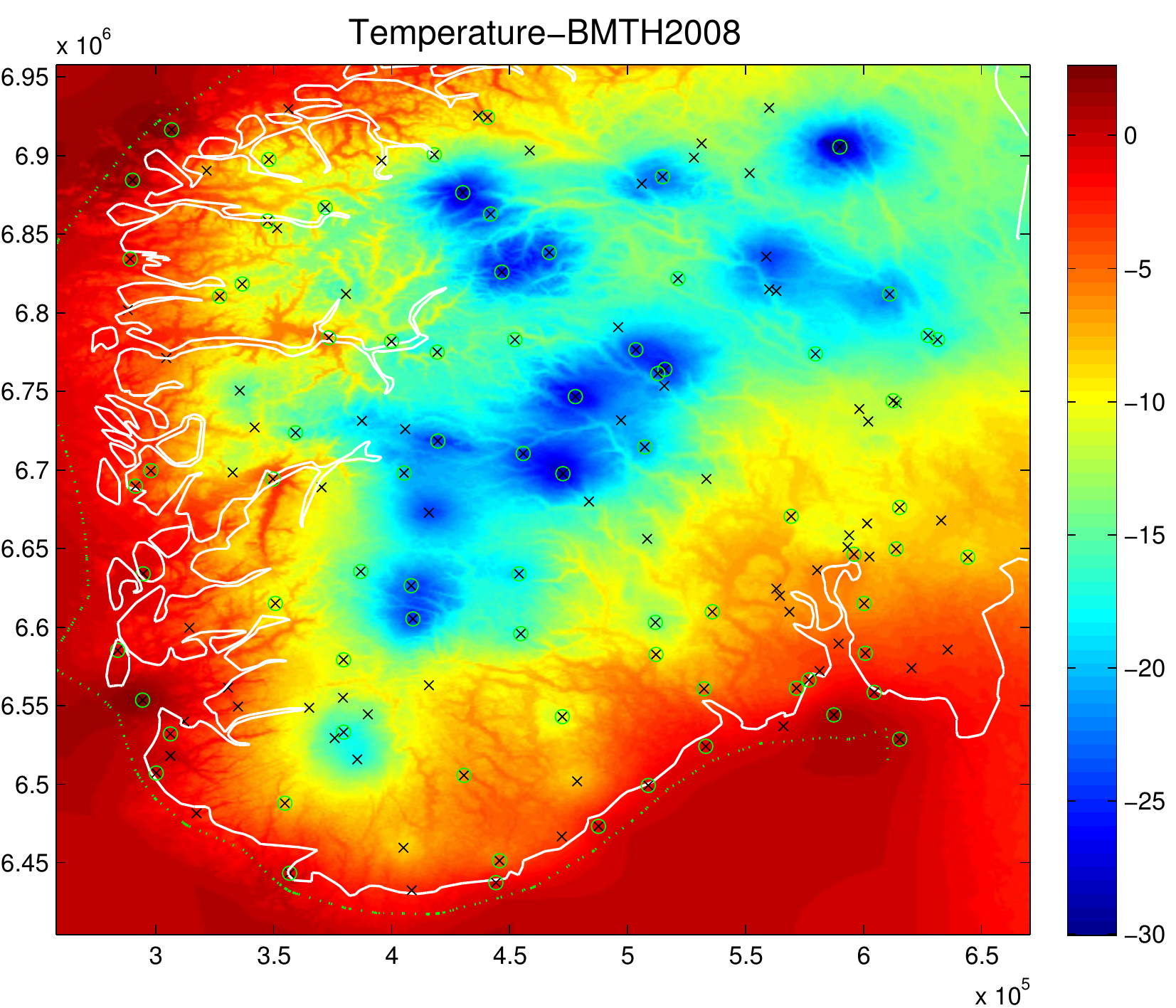} \label{fig: spdenorway_reconstruct_temp_2008BM}}
     \subfigure[]{\includegraphics[width=0.3\textwidth,height=0.3\textwidth]{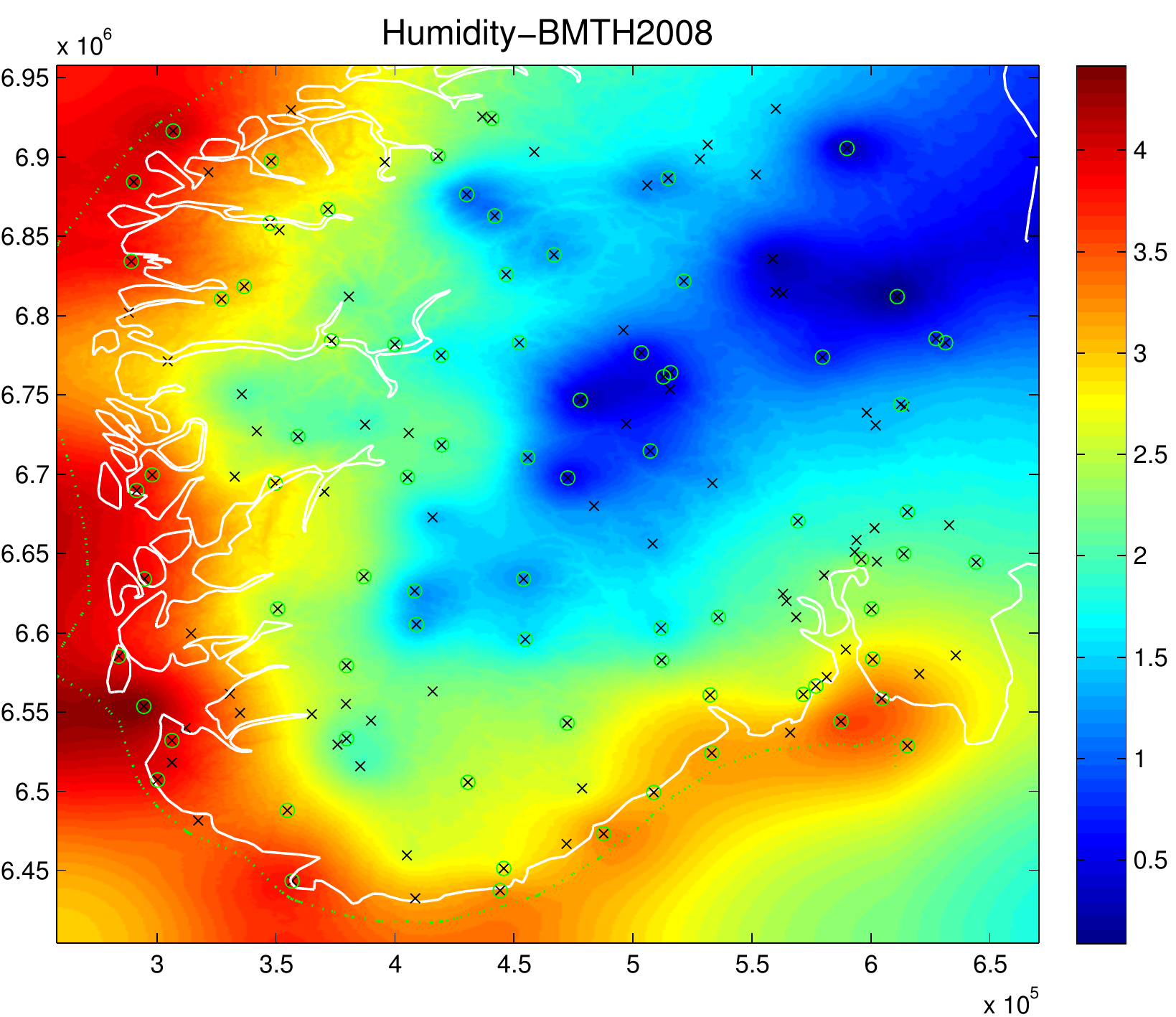} \label{fig: spdenorway_reconstruct_humi_2008BM}}\\
     \subfigure[]{\includegraphics[width=0.3\textwidth,height=0.3\textwidth]{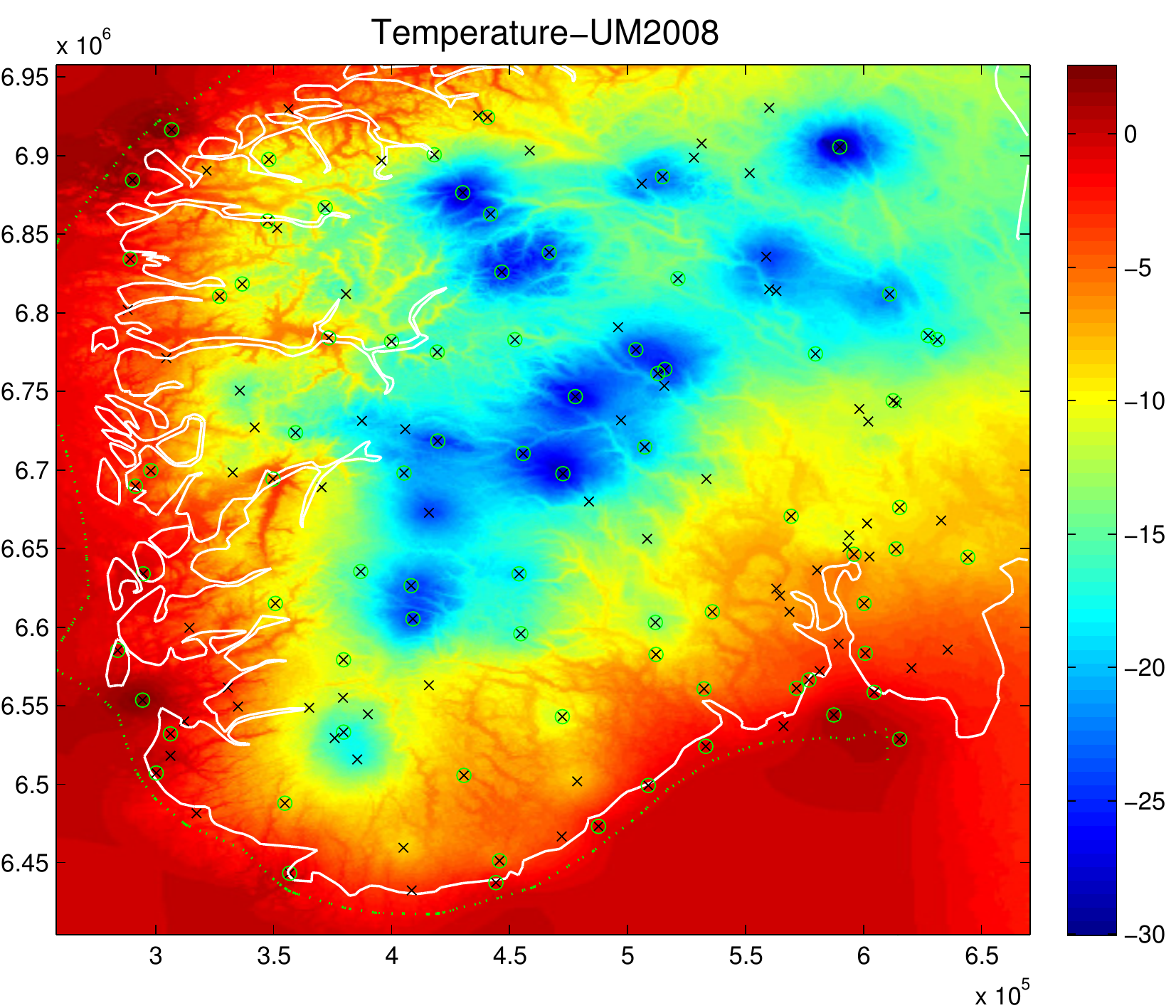} \label{fig: spdenorway_reconstruct_temp_2008UM}} 
     \subfigure[]{\includegraphics[width=0.3\textwidth,height=0.3\textwidth]{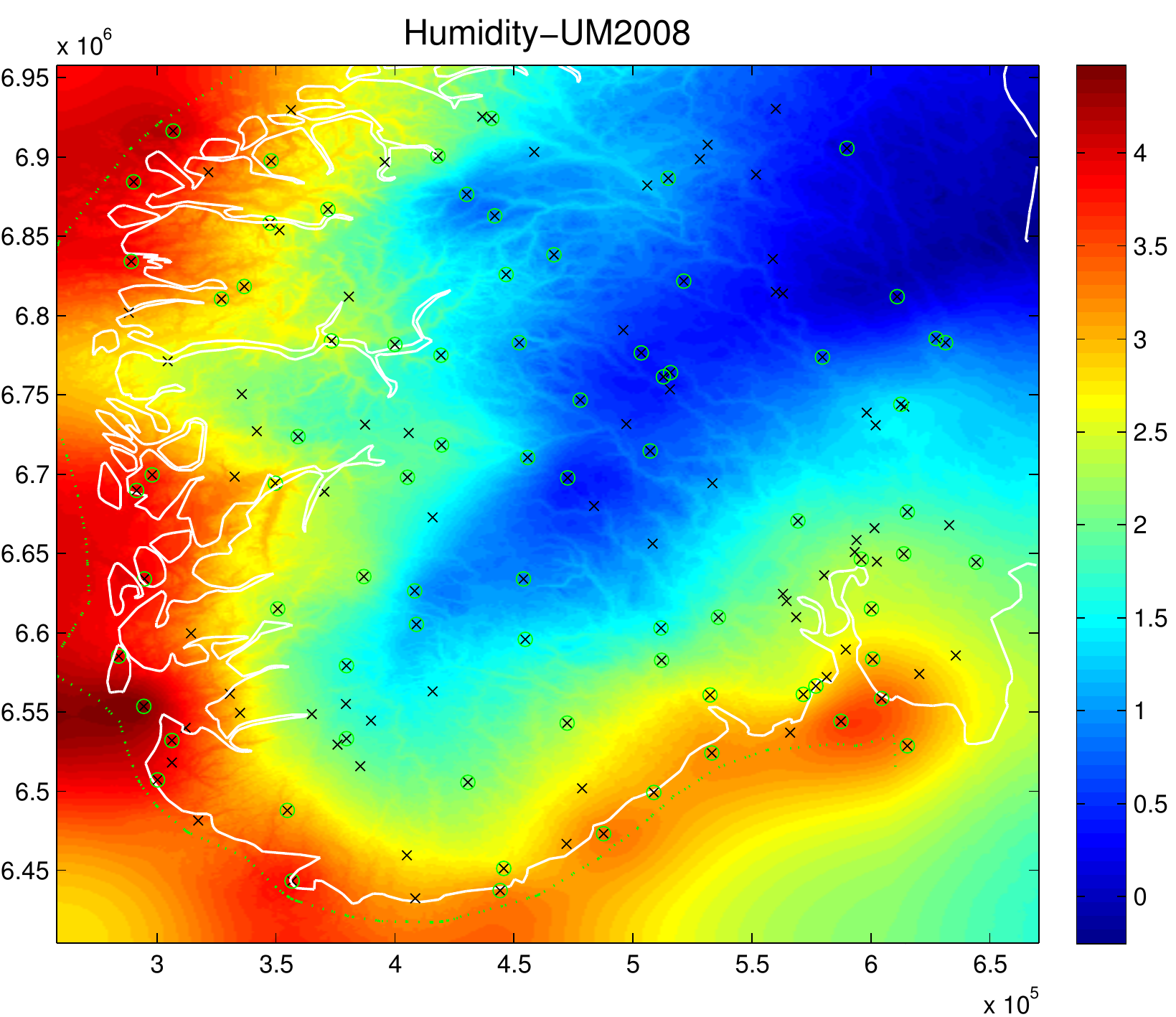} \label{fig: spdenorway_reconstruct_humi_2008UM}}\\
     \subfigure[]{\includegraphics[width=0.3\textwidth,height=0.3\textwidth]{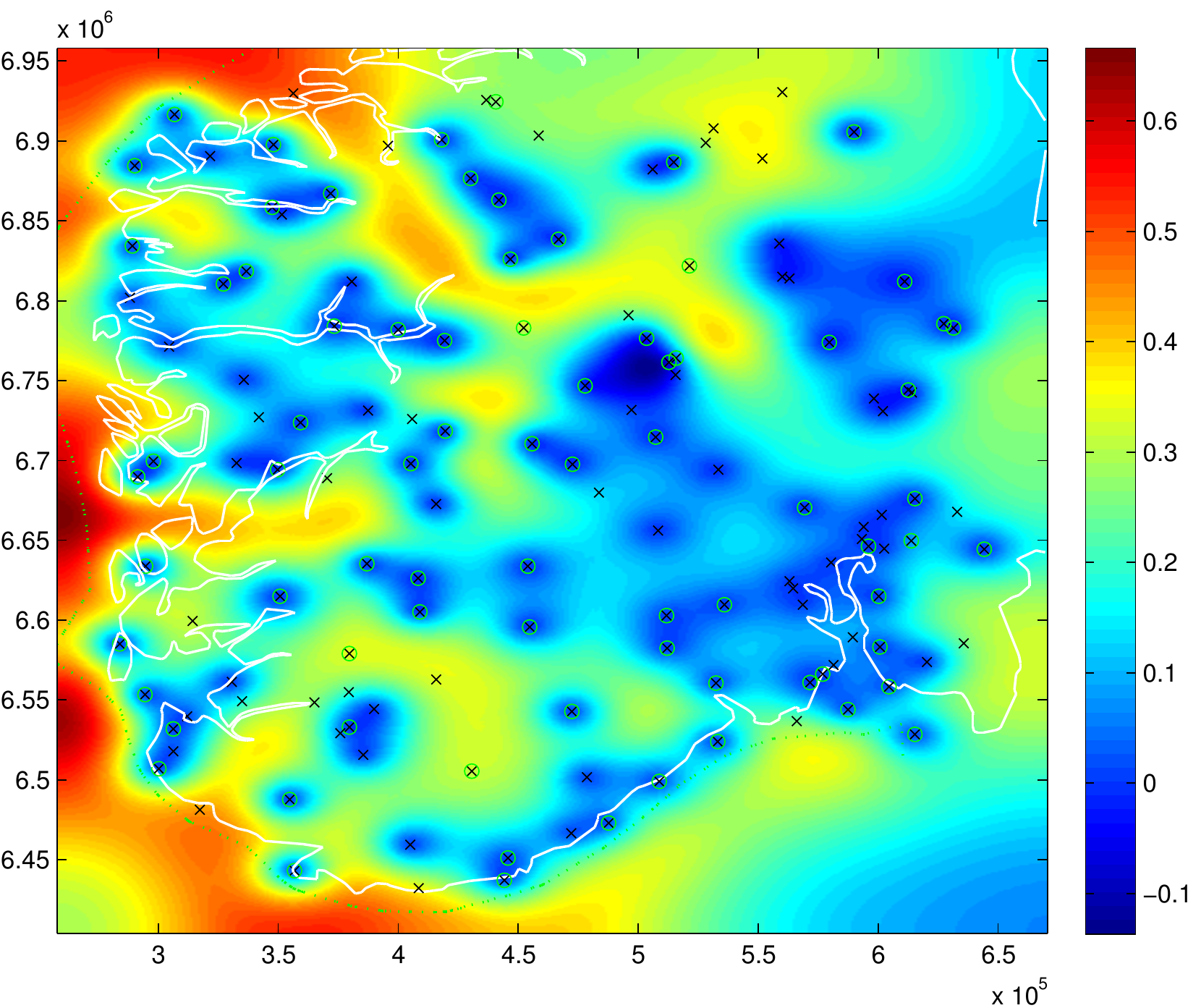} \label{fig: spdenorway_reconstruct_diffT}} 
     \subfigure[]{\includegraphics[width=0.3\textwidth,height=0.3\textwidth]{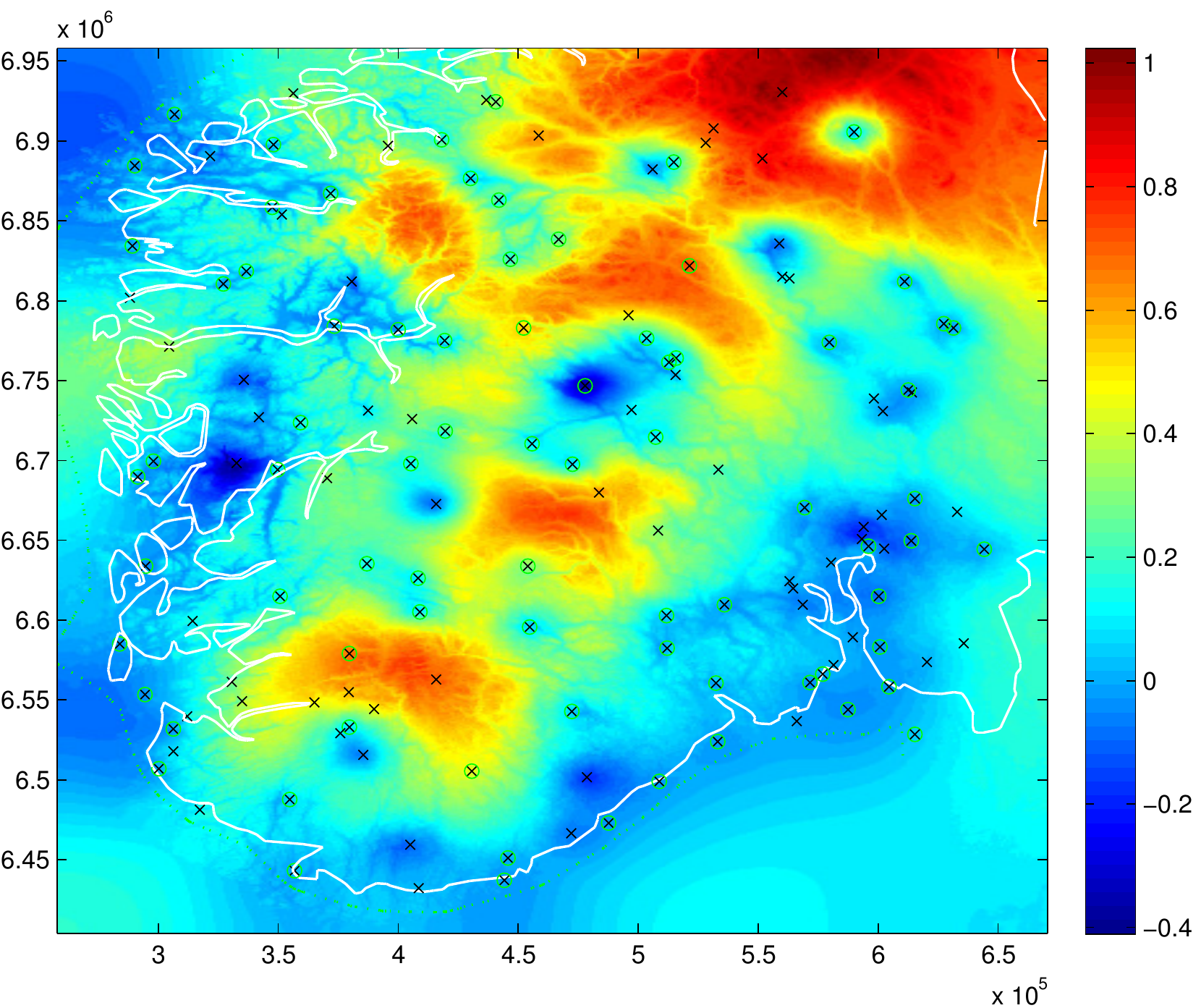} \label{fig: spdenorway_reconstruct_humi_diffH}}
    \caption{Reconstructed temperature and humidity in $2008$ for bivariate model (BMTH) (a) - (b) and 
             for univariate model (c) - (d) with $1 \text{km} \times 1 \text{km}$ resolution together with the differences (e) - (f) between these two models.
             The red cross marks ($\times$) and the green circle marks ($\circ$) are locations for temperature and humidity observations, respectively.} 
\label{fig: spdenorway_reconstruct_2008}
\end{figure}

With the estimates given in Section \ref{sec: spdenorway_bayesianparameters}, we can reconstruct temperature and humidity over 
the northern Norway with $1$km by $1$km resolution. Figure \ref{fig: spdenorway_reconstruct_2008} shows the reconstructed 
temperature and humidity in $2008$ for bivariate model (BM-TH) (a) - (b) and for univariate model (c) - (d).
The fields are reconstructed by first estimating the relevant parameters with lower resolution model, 
and then use the posterior modes for the parameters and the covariates for the $1$km by $1$km resolution model. 
The differences between these two models are shown in Figure \ref{fig: spdenorway_reconstruct_diffT} and in Figure \ref{fig: spdenorway_reconstruct_humi_diffH}.
From these two figures we find that the differences at the locations where the observations are available are small, while the differences are larger when it is further away from
the observations. With the bivariate model we use the variance-covariance structure to borrow information between humidity 
and temperature. From Figure \ref{fig: spdenorway_reconstruct_humi_diffH} we can also notice that humidity is less influenced by elevation in the bivariate model than the 
univariate mode since the bivariate model borrows some information from temperature and we have more temperature observations in the dataset.
It is further illustrated with the predictive performance in Section \ref{sec: spdenorway_predictiveperformance}.

\subsection{Predictive performance} \label{sec: spdenorway_predictiveperformance}
In this section the predictive performance for the bivariate models (BM-TH and BM-HT) and univariate model (UM) are compared
using the scores and validation scheme from Sections \ref{sec: spdenorway_scoringrules} and \ref{sec: spdenorway_crossvalidation}.

\begin{figure}[tbp]
    \centering
    \subfigure[]{\includegraphics[width=0.4\textwidth,height=0.4\textwidth]{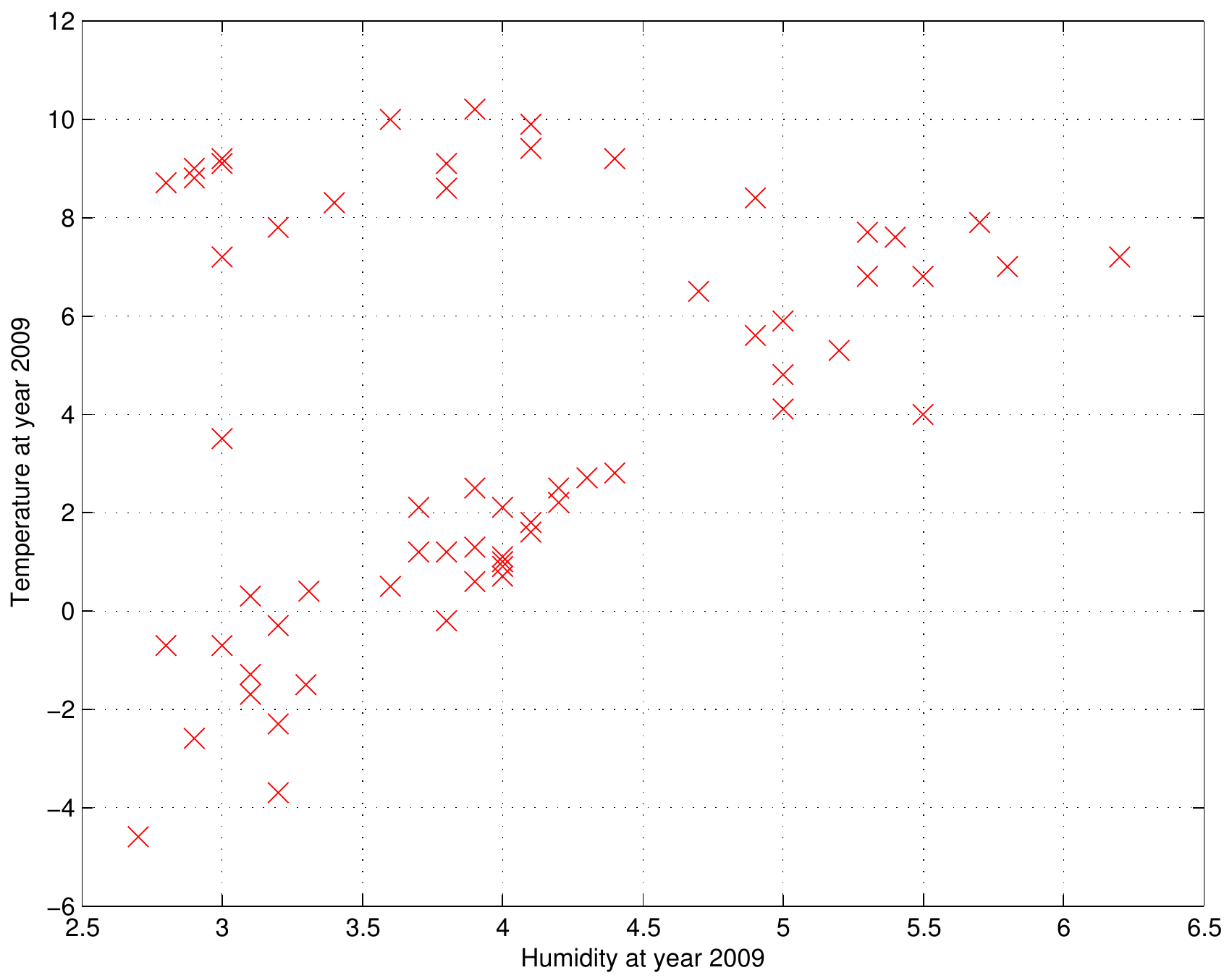}\label{fig: spdenorway_2009fulldata}}
    \subfigure[]{\includegraphics[width=0.4\textwidth,height=0.4\textwidth]{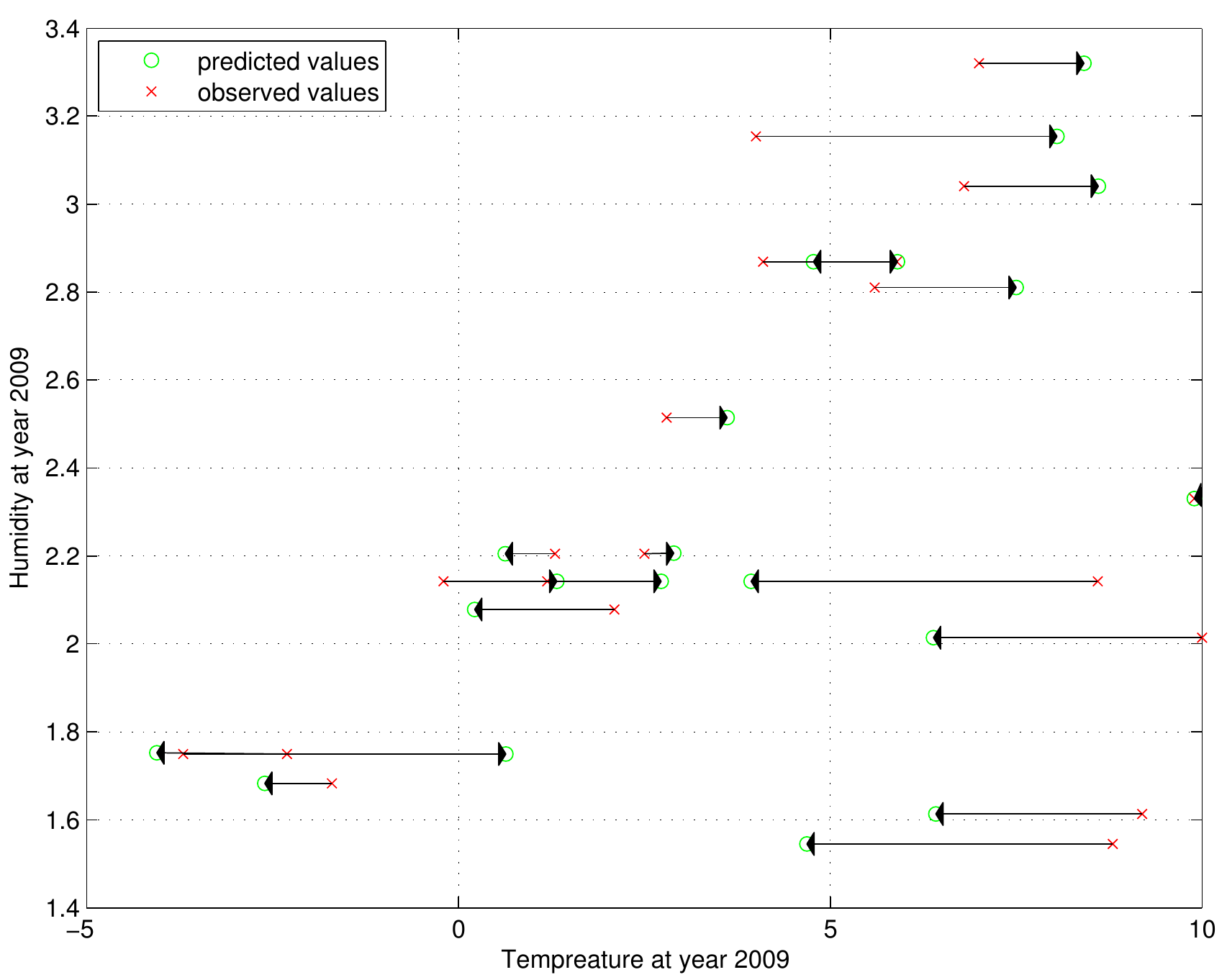} \label{fig: spdenorway_2009testdata} }
    \caption{ The dataset where both temperature and humidity are available (a) and test dataset of temperature and humidity with red crosses (b) at 2009.
              The green circles are the predicted values for humidity and temperature with model BM-TH and validation setting 'H'. The arrows connect the 
              corresponding observations and predicted values.} 
    \label{fig: spdenorway_datasets2009}
\end{figure}

From Figure \ref{fig: spdenorway_datasets2009} we can see that there are some ``outliers'' in the temperature observations in year $2009$: 
there are some locations with very high temperature but rather low humidity. This will cause poor predictive performance. 
We therefore now first discuss predictions based on results excluding 2009 from the test dataset, and then come back to
$2009$ later.
Figure \ref{fig: spdenorway_2009fulldata} illustrates the dataset where both temperature and humidity observations are available and 
Figure \ref{fig: spdenorway_2009testdata} shows the test dataset.

\begin{figure}[tbp]
    \centering
     \subfigure[]{\includegraphics[width=0.45\textwidth,height=0.4\textwidth]{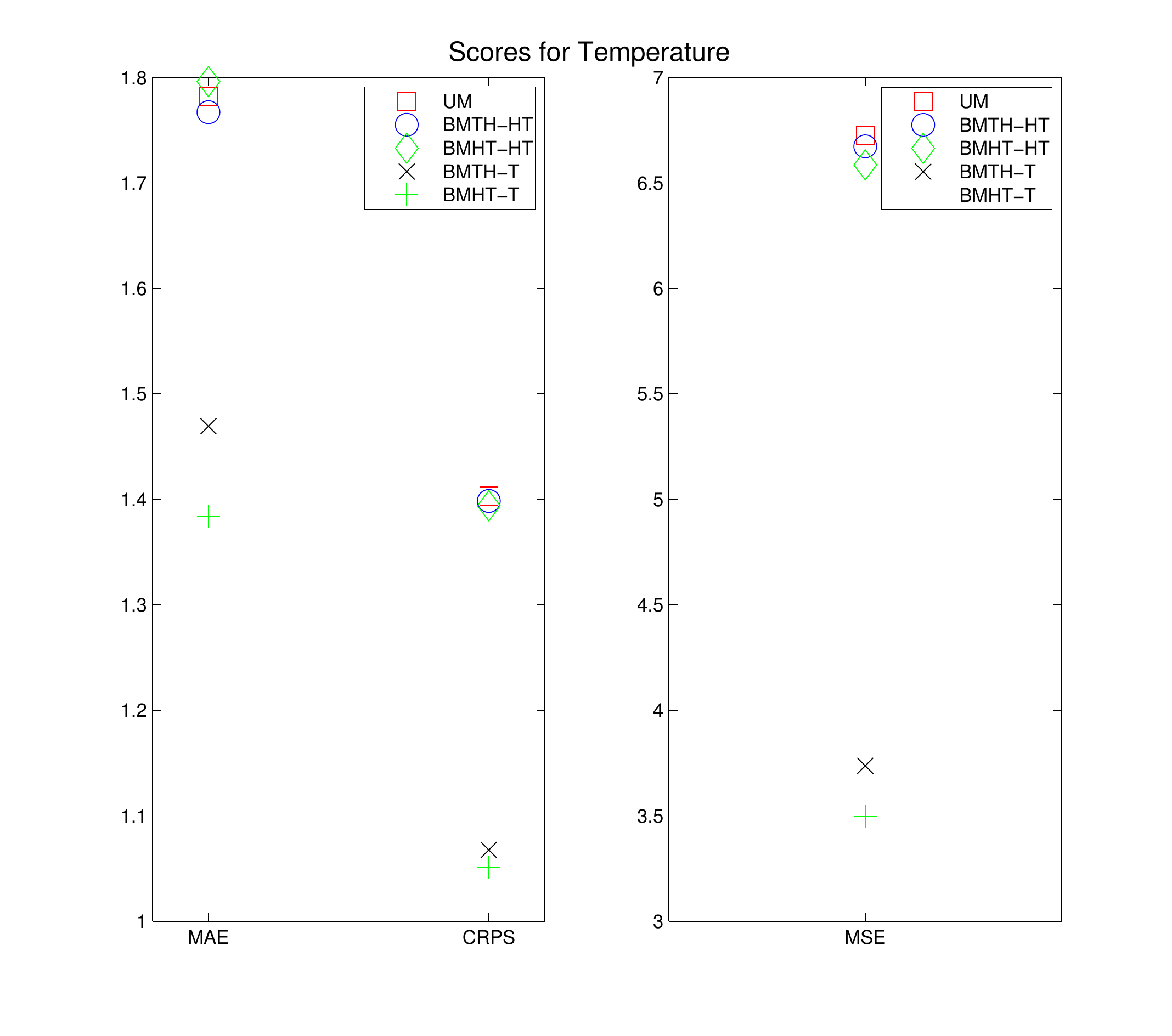} \label{fig: spdenorway_Score_T}}
     \subfigure[]{\includegraphics[width=0.45\textwidth,height=0.4\textwidth]{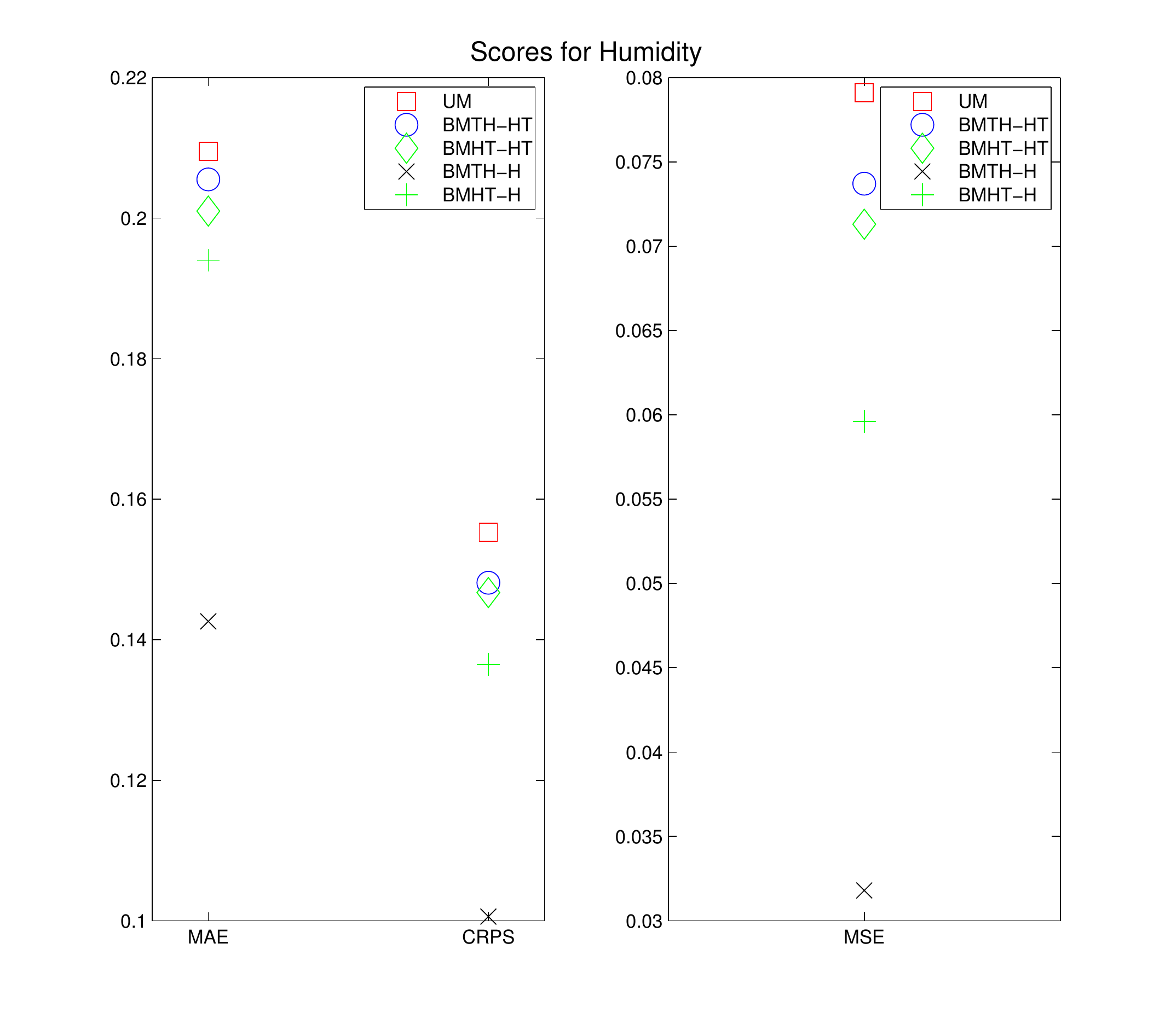} \label{fig: spdenorway_Score_H}}\\
    \caption{Predictive performance for all years except 2009 for temperature and humidity with different models and settings. } 
\label{fig: spdenorway_Score}
\end{figure}

The scores for predictions are given in Figure \ref{fig: spdenorway_Score}, and Figure \ref{fig: spdenorway_Score_T} and Figure \ref{fig: spdenorway_Score_H}
illustrate the predictive scores for temperature and humidity, respectively. In these figures, ``BMTH'' and ``BMHT''  denote the bivariate model with temperature as the 
first field and humidity as the second field, and with humidity as the first field and first as the second field, respectively. ``-T'', ``-H'' and ``-HT'' denote the validation settings.
From the results we can notice that the bivariate model with Settings ``H'' and ``T'' perform better than the univariate model for all scores.  
We can also notice that the bivariate model with Settings ``H'' and ``T'' perform better than the bivariate model with Setting ``HT''.
In addition, we can notice that the bivariate model with Setting ``HT'' performs better than the univariate models.
In other words, when observations from one field are available (validation setting ``H'' and setting ``T''), the bivariate models perform better than the univariate model. 
Further, if neither temperature nor humidity observations are available at the test locations (validation setting ``HT''), 
the bivariate models perform better than the univariate model, but not as good as when observations of the other quantity is available at the test locations.

From Figure \ref{fig: spdenorway_Score} we further notice that the order of the fields matters for the predictive performance. 
It shows that we get better results when we set the corresponding field as the second field, especially for validation settings ``H'' and ``T''.
For instance, if we are interested in predicting humidity, the result is better when it is set as the second field, 
especially when temperature observation at the test locations are available. 
However, the bivariate models perform better than the univariate model regardless of the order of fields.

Regarding year $2009$, the bivariate models perform worse than the univariate model, especially when the other quantity is available at the test location
, i.e., the CRPS values temperature with 'UM', 'BMTH' with validation setting 'HT', and
'BMTH' with validation setting 'H' are $1.15, 1.18$ and $1.46$, respectively, 
because the 'borrowed' information is wrong in the bivariate models. 
This result is useful since in this case it can be used as an indicator of outliers in our dataset which might need special treatment. 
We notice in Figure \ref{fig: spdenorway_2009testdata} that the bivariate model 'BMTH' with validation setting 'H' tries to drag the outliers back to follow the positive correlation.

\section{Discussion and Conclusion} \label{sec: spdenorway_discussion}
In this paper we have set up, fitted and evaluated models for temperature and humidity in Southern Norway based on the 
observations on $7$th of December from $2007$ to $2011$  using elevation and distance to ocean as explanatory variables. 
Three different models are compared in this paper: two bivariate models for modelling temperature and humidity jointly, and one univariate model for 
modelling them independently. 
To set up bivariate models the system of SPDEs approach proposed by \citet{hu2012multivariate} is used, 
while the corresponding univariate approach is chosen for univariate models. 
For all models the parameters for the explanatory variables agree with physical knowledge.
Further, there are spatial dependence both for
humidity and temperature, and the bivariate models 
shows also positive spatial cross-correlation. 

To compare predictive performance between the three models,
three different validation settings are used. 
We conclude that using a bivariate GRF to model temperature and humidity jointly is superior to model them 
independently using univariate GRFs, both in term of prediction accuracy (has lower RMSE), and in term of quantifying prediction uncertainty (has lower mean CRPS). 
Using a bivariate model is especially useful for predicting humidity, as it has a sparser network than temperature. 
For locations at or close to a temperature observation the bivariate model is able to utilize this information when predicting humidity. 

The results also illustrate that the order of fields is relevant from the prediction point of view 
when we use a triangular system of SPDEs for constructing a bivariate field. 
From an applied point of view, 
the results from both orders are satisfiable, and we do not need to consider both  
if the computational resources or time is limited.
We have found that if one of the quantities is our prime interest, this should be the second field which is given a model that is a mixture of Mat\'ern models.
From a modeling point of view it is interesting that it seems to be beneficial to use a mixture
of Mat\'ern models, and this model class is an interesting topic for future research. 

From our results we have learned that the bivariate models do not always perform better than the univariate model. In year $2009$
there were a group of observations that did not follow the general positive dependency between temperature and humidity, but seemed to be independent.
One way to tackle this could be to extend the bivariate model to allow for a spatial varying dependency. To set up such a model has to be done carefully
to ensure positive definite covariance functions and to keep the computational efficiency, and this is outside the scope of this paper.

There might be some other explanatory variables, such as wind speed and solar radiation, which should be included in the model. 
Further, there is, for a given pressure and temperature an upper limit of humidity \citep{barry2010atmosphere}. 
This physical limitation is not included in our model. All these might improve the predictive performance of our model. 
On the other hand, the purpose of our modeling is to provide input to a deterministic physically based model for precipitation. 
This model would convert the nonphysically high humidity to precipitation, which might give good predictions for precipitation. 
From an applied point of view, we find incorporating our results with a physical model for precipitation and evaluate the 
differences between our models with respect to precipitation predictions is the most interesting direction for further work.

\section*{Appendix A. Gaussian Markov random fields} \label{sec: appendixA}

A random vector $\boldsymbol{x} = \left( x_1, x_2, \dots, x_n \right) \in \mathbb{R}^n$ is a Gaussian random field with mean $\boldsymbol{\mu}$ and
precision matrix $\boldsymbol{Q} > 0$ ($\boldsymbol{Q} = \boldsymbol{\Sigma}^{-1}$) if and only if its density is
\begin{equation}
 \pi(\boldsymbol{x}) = \frac{1}{(2\pi)^{n/2}}|\boldsymbol{Q}|^{1/2}\exp\left({\frac{1}{2}(\boldsymbol{x}-\boldsymbol{\mu})^{\mbox{T}}\boldsymbol{Q}(\boldsymbol{x}-\boldsymbol{\mu})}\right).
 \label{eq: spdenorway_GMRFdensity}
\end{equation}
where $\boldsymbol{x}_{-ij}$ denotes for $\boldsymbol{x}_{-\{i,j\} }$. $\boldsymbol{Q} > 0$ denotes that it is positive definite. 
Gaussian \emph{Markov} random fields are the main tool for achieving computational efficiency with models built by the SPDE approach.
A Gaussian Markov random fields is a GRF with \emph{Markov} property
\begin{equation}
 Q_{ij} = 0  \Longleftrightarrow  x_i \perp x_j|\boldsymbol{x}_{-ij},
\end{equation}

\noindent and hence the precision matrix $\boldsymbol{Q}$ for a GMRF is usually sparse. Therefore, numerical algorithms 
for sparse matrices can be applied when doing computations. 
\citet{rue2005gaussian} gives a more detailed discussion on the theories for GMRFs. A condensed discussion about GMRFs can also be found 
in \citet[Chapter $12$]{gelfand2010handbook}. 

\section*{Appendix B. Inference} \label{sec: appendixB}

Since the coefficient parameters for the covariates can be modelled with Gaussian distributions, 
we can treat the coefficients $\boldsymbol{\beta}_{j}$ as part of the latent field together the spatial process $\boldsymbol{x}(\boldsymbol{s})$ 
and model them jointly instead of treating the coefficient parameters
as hyper-parameters. The hyper-parameters then only contains the 
parameters from the systems of SPDEs \eqref{eq: spdenorway_SPDEs_system_triangular},
$\boldsymbol{\theta} = \left\{ b_{11}, b_{21}, b_{22}, \kappa_{11}, \kappa_{21}, \kappa_{22} \right\}$ 
for bivariate model and 
$\boldsymbol{\theta} = \left\{ b_{11}, b_{22}, \kappa_{11}, \kappa_{22} \right\}$ for univariate model, 
since we fix the values of $\{\alpha_{ij}; i, j = 1,2\}$ for both the models.
The latent field in this case is $\boldsymbol{z} = \left( \boldsymbol{x}, \boldsymbol{\beta} \right)^{\mbox{T}}$, 
where $\mbox{T}$ denotes the transpose of a vector or a matrix.
This can speed up the optimization considerably since there are much fewer parameters in the numerical optimization. 
This is the commonly used setting in \citet{rue2009approximate}.

Let $\boldsymbol{Q}(\boldsymbol{\theta})$ denote
the precision matrix for the random fields constructed by the system of SPDEs \eqref{eq: spdenorway_SPDEs_system_triangular}
for the bivariate GRFs or the precision matrix for the univariate random fields with 
SPDE \eqref{eq: spdenorway_spde_simple} with hyper-parameters $\boldsymbol{\theta}$. 
With the univariate model we construct the precision matrix $\boldsymbol{Q}(\boldsymbol{\theta})$ as a block diagonal precision matrix, 
then inference for this two univariate random fields can be done simultaneously. 
In this case we can use the same program for the bivariate model, and the univariate model has only one more constraint $b_{21} = 0$.
\citet{hu2012multivariate} have shown that from the well known Bayesian formula 
\begin{equation} \label{eq: spdenorway_bayesianformula}
 \pi(\boldsymbol{y}, \boldsymbol{\theta}) 
  = \frac{\pi \left( \boldsymbol{\theta}, \boldsymbol{z}, \boldsymbol{y} \right)}{\pi \left(\boldsymbol{z}|\boldsymbol{y}, \boldsymbol{\theta} \right)},
\end{equation}
we can derive the posterior distribution

\begin{equation}
  \begin{split}
 \log \left( \pi\left(  \boldsymbol{\theta} |\boldsymbol{y} \right) \right) 
 = & \text{ Const.} + \log \left( \pi \left( \boldsymbol{\theta} \right)  \right) + \frac{1}{2}\log \left( |\boldsymbol{Q}(\boldsymbol{\theta})| \right) \\
   & - \frac{1}{2}\log \left( |\boldsymbol{Q}_c(\boldsymbol{\theta})| \right) 
   + \frac{1}{2} \boldsymbol{\mu}_c^{\mbox{T}}(\boldsymbol{\theta})\boldsymbol{Q}_c(\boldsymbol{\theta}) \boldsymbol{\mu}_c(\boldsymbol{\theta}),
  \end{split}
\label{eq: spdenorway_posterialformula}
\end{equation}
with $\boldsymbol{\mu}_c = \boldsymbol{Q}_c^{-1}\boldsymbol{C}^{\mbox{T}}\boldsymbol{Q}_{\epsilon}\boldsymbol{y}$, 
~$\boldsymbol{Q}_c(\boldsymbol{\theta}) = \boldsymbol{Q}(\boldsymbol{\theta}) + \boldsymbol{C}^{\mbox{T}}\boldsymbol{Q}_{\epsilon}\boldsymbol{C}$, and
$\boldsymbol{C} = \left(\boldsymbol{A}, \boldsymbol{X}\right)$.
$\boldsymbol{A}$ is a sparse matrix which links the sparse observations of temperature and humidity to our bivariate GRF or univariate GRFs. 
$\boldsymbol{X}$ is the design matrix.

\bibliographystyle{plainnat}
\bibliography{Ref.bib}

\end{document}